\documentclass{aastex}
\usepackage{emulateapj5}
\newcommand{\chandra}{{\itshape Chandra}}
\newcommand{\lum}{erg s$^{-1}$ }
\newcommand{\flux}{erg cm$^{-2}$ s$^{-1}$ }
\newcommand{\enedens}{erg cm$^{-3}$}
\newcommand{\colden}{cm$^{-2}$}
\newcommand{\ue}{$u_{\rm e}$}
\newcommand{\um}{$u_{\rm m}$}
\slugcomment{}
\shorttitle{Diffuse hard X-ray emission associated with the lobes of 3C~452.}
\shortauthors{N. Isobe et al.}

\begin{document}
\title{A \chandra~detection of diffuse hard X-ray emission 
associated with the lobes of the radio galaxy 3C~452}
\author{N. Isobe\altaffilmark{1,3},
	M. Tashiro\altaffilmark{2}. 
	K. Makishima\altaffilmark{3,4},
	N. Iyomoto\altaffilmark{5},
	M. Suzuki\altaffilmark{2},
	M. M. Murakami\altaffilmark{3},
	M. Mori\altaffilmark{2,4},
	and 
	K. Abe\altaffilmark{2}
	}
\email{isobe@oasis.tksc.nasda.go.jp}
\altaffiltext{1}{Tsukuba Space Center,
National Space Development Agency of Japan,
	2-1-1, Sengen, Tsukuba, Ibaraki, 305-8505, Japan.}
\altaffiltext{2}{Department of Physics, Saitama University,
	Shimo-Okubo, Saitama, 338-8570, Japan.}
\altaffiltext{3}{Department of Physics, University of Tokyo,
	7-3-1, Hongo, Bunkyo, 113-0033, Japan.}
\altaffiltext{4}{Cosmic Radiation Laboratory, 
	the Institute of Physical and Chemical Research,
	2-1 Hirosawa, Wako, Saitama, 351-0198, Japan}
\altaffiltext{5}{The Institute of Space and Astronautical Science, 
	3-1-1 Yoshinodai, Sagamihara, Kanagawa, 229-8510, JAPAN}
\begin{abstract}
An 80 ksec \chandra~ACIS observation
of the radio galaxy 3C~452 is reported.
A diffuse X-ray emission associated with the lobes 
has been detected with high statistical significance, 
together with the X-ray nucleus of the host galaxy. 
The 0.5--5 keV ACIS spectrum of the diffuse emission is 
described by a two-component model,
consisting of a soft thermal plasma emission 
from the host galaxy halo and 
a hard non-thermal power-law component.
The hard component is ascribed to the inverse Comptonization 
of cosmic microwave background photons by  
the synchrotron emitting electrons in the lobes,
because its spectral energy index, $0.68\pm0.28$, 
is consistent with the radio synchrotron index, $0.78$.
These results reveal a significant electron dominance in the lobes.
The electrons are inferred to have a relatively uniform distribution,
while the magnetic field is compressed toward the lobe periphery. 
\end{abstract}


\section{Introduction}
Relative dominance of electrons and magnetic fields 
in lobes of radio galaxies provides 
one of the important clues to 
the formation of astrophysical jets and evolution of radio galaxies.
The electrons in the lobes,  
while emitting powerful synchrotron radiation (SR),
can boost up seed soft photons via inverse-Compton (IC) process  
into X-rays and $\gamma$-rays,
which take the same slope as the SR radio spectrum.
Once the source of the seed photons are identified,
a comparison of the SR and the IC fluxes from the lobes 
allows us to independently determine 
energy densities of the electrons and magnetic field, 
\ue~and \um, respectively.

 These IC X-rays have been detected with {\it ASCA} and {\it ROSAT} 
from the lobes of several radio galaxies,
such as Fornax A (Feigelson et al. 1995; Kaneda et al., 1995) and 
Centaurus B (Tashiro et al. 1998)
where the cosmic microwave background (CMB) photons are boosted up,   
and 3C 219 (Brunetti et al. 1999) where  the infra-red (IR) radiation 
from the active nucleus provides the dominant seed photons. 
Although a novel possibility of significant electron dominance 
has been reported for Centaurus B and 3C 219,
the sample remained too small 
because only a few lobes are larger than the angular resolution of {\it ASCA}. 

Obviously, 
the \chandra~ACIS is expected to innovate the study of IC X-rays 
from a large sample of radio lobes.   
Actually, the IC X-rays are being detected by the ACIS 
from the lobes of 
a few objects (e.g Brunetti et al., 2001) 
and from a number of radio hot spots (e.g. Wilson et al., 2000).  
Here, we report on our \chandra~ observation of 3C~452,
in which IC emission from 
its relaxed lobes has been detected with unprecedented significance.

Throughout the present letter, 
we adopt a Hubble constant of $H_{0}= 75$ km sec$^{-1}$ Mpc$^{-1}$ and 
a deceleration parameter of $q_{0} = 0.5$. 
Hence, $1'$ corresponds to 82.3 kpc 
at the redshift, $z = 0.0811$ (Spinrad et al., 1985),  of 3C~452.

\section{Observation}
3C~452 is an FR II (Fanaroff, \& Riley, 1974) radio galaxy,
with an elliptical host.
Its radio images (e.g. Black et al., 1992) 
reveal a symmetrical double-lobe morphology, 
which has a relatively uniform surface brightness distribution.
It dose not lie in a rich cluster environment, 
and is hence free from thermal X-ray emission.
The total angular extent ($\sim1' \times 4'$) is very adequate 
for both the ACIS angular resolution and the field of view.
It has a moderately high radio flux density of $S_{\rm SR} = 10.5$ Jy
at 1.4 GHz (Laing, \& Peacock, 1980), 
with an accurately determined spectral index of 
$\alpha_{\rm SR} = 0.78$ between 178 and 750 MHz 
(Laing, Riley, \& Longair 1983). 
These make this object very suitable for our purpose
of searching for the diffuse IC X-rays from the lobes. 
Besides, a sign of extended weak X-ray emission from  3C~452
has already been obtained with {\it Einstein} (Feigelson, \& Berg,  1983)
with a 0.5 -- 3~keV flux of 
$F_{\rm X}\lesssim 5 \times 10^{-13}$ \flux (Fabbiano et al., 1984).
However, neither its X-ray morphology nor the spectrum 
have been determined accurately.

We performed an 80 ksec \chandra~ACIS 
observation of 3C~452 on 2001 August 25.
We placed its radio nucleus $2'$ from the nominal aimpoint 
on the ACIS-S3,
in order to cover its whole radio structure with this chip.
The data were read out with a nominal frame time of 3.2~s, 
using the {\bf Faint} format.  
Utilizing the CIAO 2.2.1 and the CALDB 2.15, 
we reprocessed the data in the standard manner.
We rejected the data obtained 
when the count rate in the source-free region 
exceeds 120\% of that averaged during the observation. 
These yielded 78.3 ksec of good exposure. 

\section{Results} 
\subsection{X-ray Image}	
\begin{figure}[htbp]
\plotone{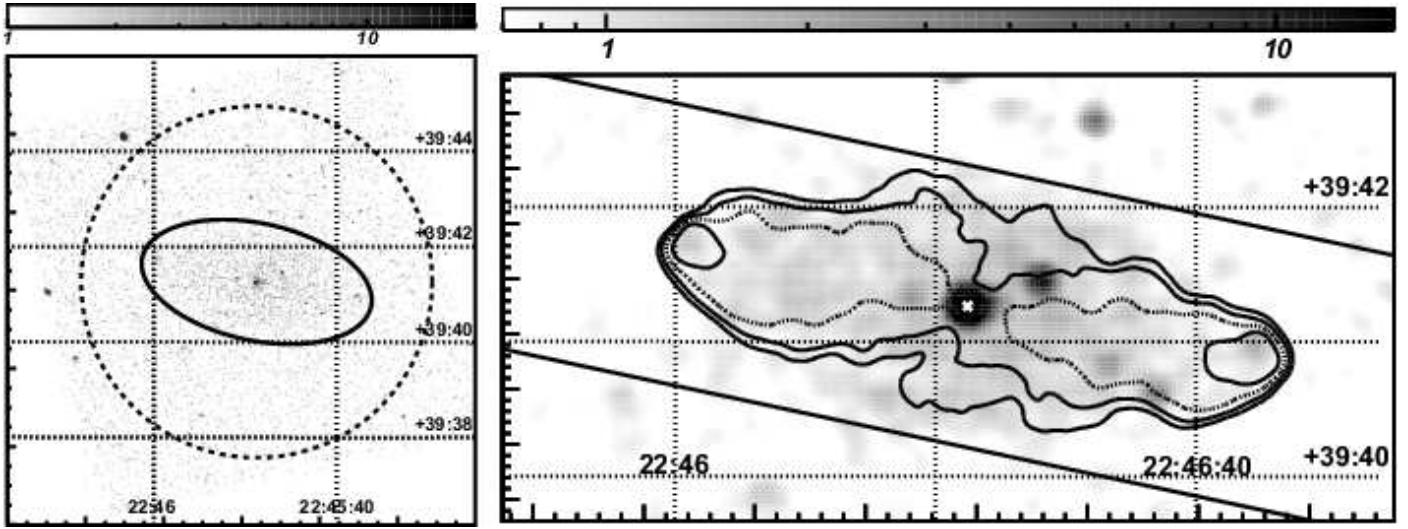}
\caption{
({\it left}) The raw ACIS-S3 image in the 0.3 -- 7 keV range,
binned into $4\times4$ ACIS pixels. 
The solid ellipse and dashed circle indicate 
integration regions for the lobe spectrum and for  
the background, respectively.      
({\it right}) The 0.3 -- 7 keV ACIS-S3 image around 3C~452 
smoothed with a two-dimensional Gaussian of $\sigma = 8$ pixels.
The 1.4 GHz VLA contours (Laing, unpublished) are overlaid.
The position of the nucleus is shown with the white cross.
The two solid lines represent the region within which the linear 
profiles are extracted. }
\label{fig:image}
\end{figure}

We show a background-inclusive 0.3 -- 7 keV ACIS-S3 image
in Figure \ref{fig:image} (left).
Among many point sources, 
one near the image center coincides with the radio nucleus. 
Moreover, we find a faint X-ray emission around the nucleus, 
extending toward the east-west direction. 
The right panel of Figure \ref{fig:image} shows the same ACIS image, 
but heavily smoothed with a two-dimensional Gaussian
of 8 ACIS pixels, i.e. $\sim4''$.
The 1.4 GHz VLA contours (Laing; unpublished) are superposed. 
The extended X-ray emission is more clearly revealed in this image;
its diffuse nature and association with the lobes are clearly seen. 

We projected the raw ACIS image 
onto the major axis of the radio lobes,
and subtracted the background
using a neighboring source free region.   
The X-ray linear surface brightness profile 
thus obtained is shown in Figure \ref{fig:profile} (a),
together with the radio profile  in Figure \ref{fig:profile} (b).
The X-ray and radio profiles have very similar overall extent,
but different structure.
The X-ray profile exhibits a center-filled morphology
with a relatively uniform surface brightness, 
while a rim-brightening feature is seen in the radio one.

\subsection{X-ray Spectrum of the nucleus}
We accumulated an ACIS spectrum within a circular region 
of $5''$ radius ($\sim 7$~kpc) centered on the nucleus.
The background spectrum was accumulated over a concentric annulus
with inner and outer radii of $6''$ and $9''$, respectively.      
Figure \ref{fig:nucl_spec} shows the background subtracted ACIS spectrum 
of the host galaxy.
It seems typical of 
so-called Compton-thick objects (e.g. Guainazzi, et al. 1999).   

We fitted the spectrum by a model consisting of 
the following 4 spectral components;
(1) a heavily absorbed power-law (PL),
(2) a Gaussian centered at $\sim6$ keV,
(3) a reflection component from neutral material 
	({\bf pexrav} with a parameter ${\rm rel\_refl} = -1$ in XSPEC;
	Magdziarz \& Zdziarski, 1995),
and 
(4) a Raymond-Smith (RS) model,
	to model a contamination 
	from thermal plasma associated with the host galaxy. 
All of them are modified by the Galactic absorption 
($N_{\rm H} = 1.2 \times10^{21}$ \colden; Dickey, \& Lockman, 1990).
We fixed the metalicity of (3) and (4) at the 0.4 solar abundance
(e.g., Matsushita, Ohashi, \& Makishima, 2000). 
We adopted a common energy index for (1) and 
the illuminating source for (2).
We assumed an inclination of $\cos \theta = 0.45$ for 
the reflecting material.
The fit has become acceptable ($\chi^2/d.o.f = 86.2/ 96$), 
with the spectral parameters summarized in Table \ref{table:nucle}.  

The obtained parameters are consistent with the view 
that a medium luminosity radio nucleus 
(Sambruna et al., 1999)
is seen through a thick material, 
which absorbs the direct continuum, 
and produces the reflection component and 
the fluorescent Fe-K$\alpha$ line.
Considering the symmetrical lobe morphology, 
the obscuring material is most likely the molecular torus 
fueling the nucleus.

\subsection{X-ray Spectrum of the lobes}
To  examine the diffuse lobe X-ray emission,
we accumulated another ACIS spectrum within a solid ellipse shown in
the left panel of Figure \ref{fig:image}, 
which contains the whole radio structure.
We rejected pixels within $6''$ of the nucleus and
those within $3''$ of possible point sources 
(defined as $\ge 10$ counts per binned pixel).
We extracted ACIS background spectra from several source-free regions 
which have the same area as the signal integration region.
Since they agreed 
to each other within $\sim 5$ \% over the 0.3 -- 7 keV range,
we have accumulated the background spectrum over 
a dashed circle of the left panel of Figure \ref{fig:image},
excluding the lobe region.
We rejected point sources 
by the same criterion as for the signal spectrum. 

Figure \ref{fig:lobe_spec} shows the background subtracted 
ACIS spectrum of the lobes of 3C~452 in the 0.5 -- 5 keV range,
where the diffuse emission is most significant (${\rm S}/{\rm N} = 38.6$). 
It appears relatively featureless and hard.
We tried to fit the data by a single PL model 
or a thermal bremsstrahlung (Bremss) model,
modified by a free absorption.
The fit was unacceptable, 
because of the data excess around 1 keV.
Both the PL and the Bremss fits have become acceptable,
by adding a soft RS component with the abundance fixed at 0.4 solar.
The obtained parameters are summarized 
in Table \ref{table:lobe_param},
and the PL + RS fit is shown in Figure \ref{fig:lobe_spec}.

\section{Discussion}
In our 80 ksec \chandra~observation of 3C~452,
we have detected the diffuse X-ray emission 
closely associated with its radio lobes.
The spectrum is successfully reproduced by the two-component model 
consisting of soft and hard components.
The absorption column density
becomes consistent with the Galactic value.
Considering the temperature and the luminosity 
in terms of the RS fit (Table \ref{table:lobe_param}), 
the soft component is naturally attributed to thermal emission
from the hot halo of the host galaxy  
(e.g., Matsushita, Ohashi, \& Makishima, 2000).

If we adopt the Bremss model for the hard component,
the temperature, $kT\gtrsim6$~keV, 
and the 2--10 keV luminosity,  
$L_{\rm X} \sim 3.3 \times 10^{42} $ \lum, 
are both rather high compared with those of thermal plasma 
of nearby elliptical galaxies.
Moreover, the thermal pressure, $p \sim 10^{-11} $ dyne cm$^{-2}$,
is nearly two orders of magnitude larger 
than the minimum-energy non-thermal pressure of the lobes,
$p_{\rm me} \sim 10 \time 10^{-13}$ dyne cm$^{-2}$.
We regard these situations as unrealistic.
On the other hand, 
the PL modeling of the hard component gives a  
spectral index, $\alpha_{\rm X} = 0.68 \pm 0.28$, 
which agrees with the SR index of the lobes,  
$\alpha_{\rm SR} = 0.78 $.
We therefore conclude that 
the diffuse hard X-rays are produced via the IC process
by the SR electron in the lobes. 

Among several candidates for the seed photon source, 
the IR photons from the active nucleus are expected to 
provide an energy density $ \lesssim 10^{-14}$ \enedens,
because the nucleus is not luminous (Lilly, Longair, and Miller 1985).
We estimate the SR photons to have an energy density 
$\lesssim 10^{-16}$ \enedens, using the lobe SR spectrum and image.
Since these energy densities fall much below that of the CMB,
$5.6\times10^{-13}$ \enedens~at the redshift of 3C~452,
we conclude that the CMB provides the seed photons.

Let us evaluate \ue~and \um, spatially averaged over the lobes.
We assume that the electron number spectrum is a single PL 
with a spectral index $p=2\alpha_{\rm SR} + 1$.
The upper and lower limits of the electron Lorentz factor
are assumed to be $10^3$ and $10^5$, respectively,
in order to cover all of the observable SR and IC frequencies.
We also assume that 
the magnetic field is randomly oriented 
and the electron velocity is isotropically distributed.
According to Harris \& Grindlay (1979), 
a comparison of the SR and the IC flux densities 
allows us to derive  
	$u_{\rm e} = (2.1\pm0.7) \times 10^{-12}$ erg cm$^{-3}$
and 
	$u_{\rm m} = 7.7_{-2.3}^{+4.5} \times 10^{-14}$ erg cm$^{-3}$; 
the latter gives a  magnetic field strength of 
$B = 1.4_{-0.3}^{+0.4}$~$\mu$G. 
Thus, we find that a significant electron dominance
is achieved in the lobes, 
as represented by the ratio $u_{\rm e}/u_{\rm m} = 27_{-16}^{+25}$.

\begin{figure}[htbp]
\plotone{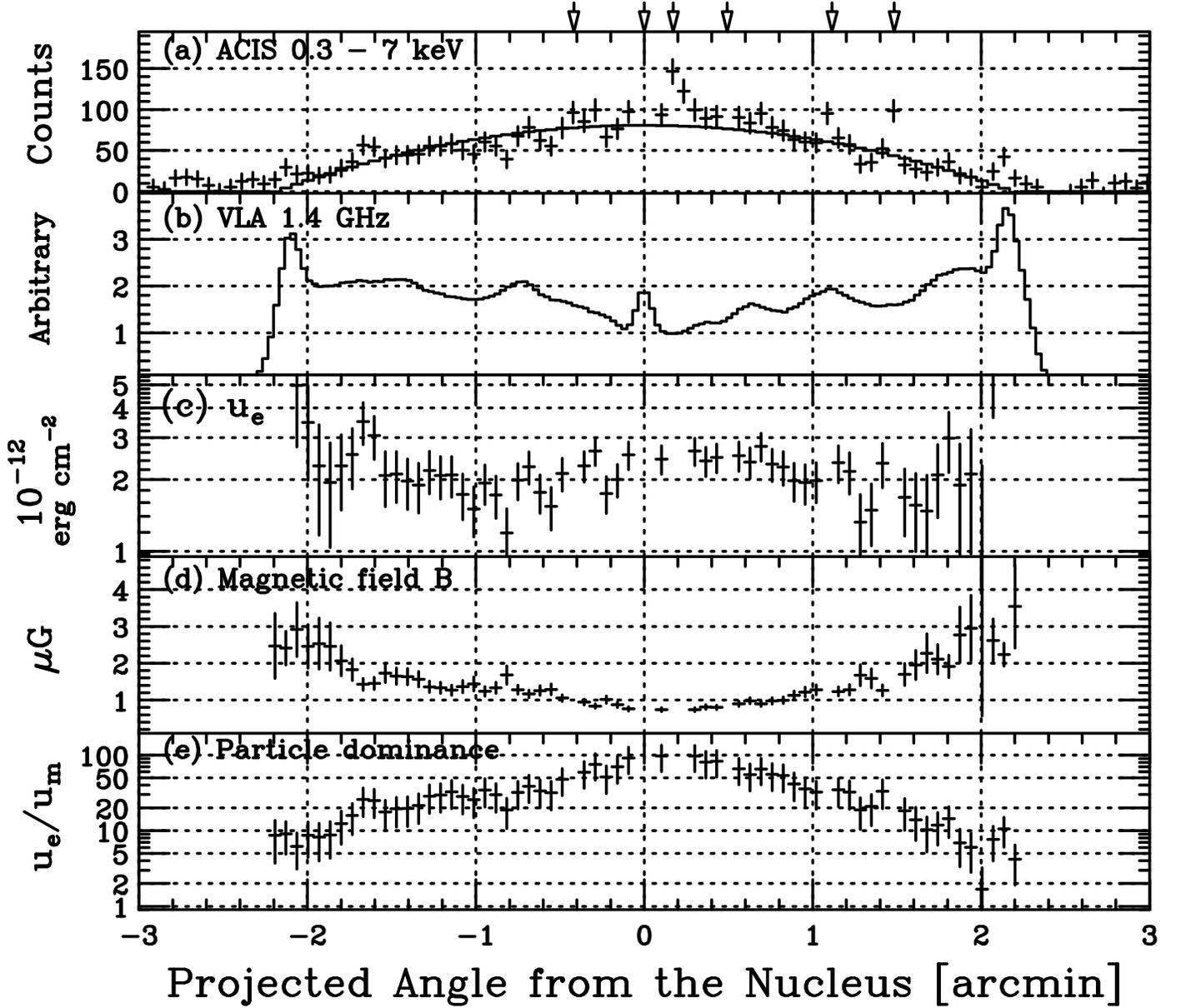}
\caption{The linear profiles 
along the major axis of the lobes of 3C~452,
of (a) the X-rays in the 0.3 -- 7 keV range,  
(b) the SR radio emission at 1.4 GHz,
(c) the electron energy density \ue,
(d) the magnetic fields $B$,
and  
(e) the ratio $u_{\rm e}/u_{\rm m}$.
The data were accumulated between the two solid lines 
in Figure \ref{fig:image} (right).
The X-ray data point for the nucleus at angle = 0
is truncated. 
Arrows on the top of the figures indicate 
the projected position of some point sources,
and data points corresponding to these arrows and to the nucleus  
are rejected in panels (c), (d), and (e). 
The histogram in panel (a) represents the model prediction 
for the IC X-ray brightness profile discussed in \S 4. 
}
\label{fig:profile}
\end{figure}

\begin{figure}[htbp]
\plotone{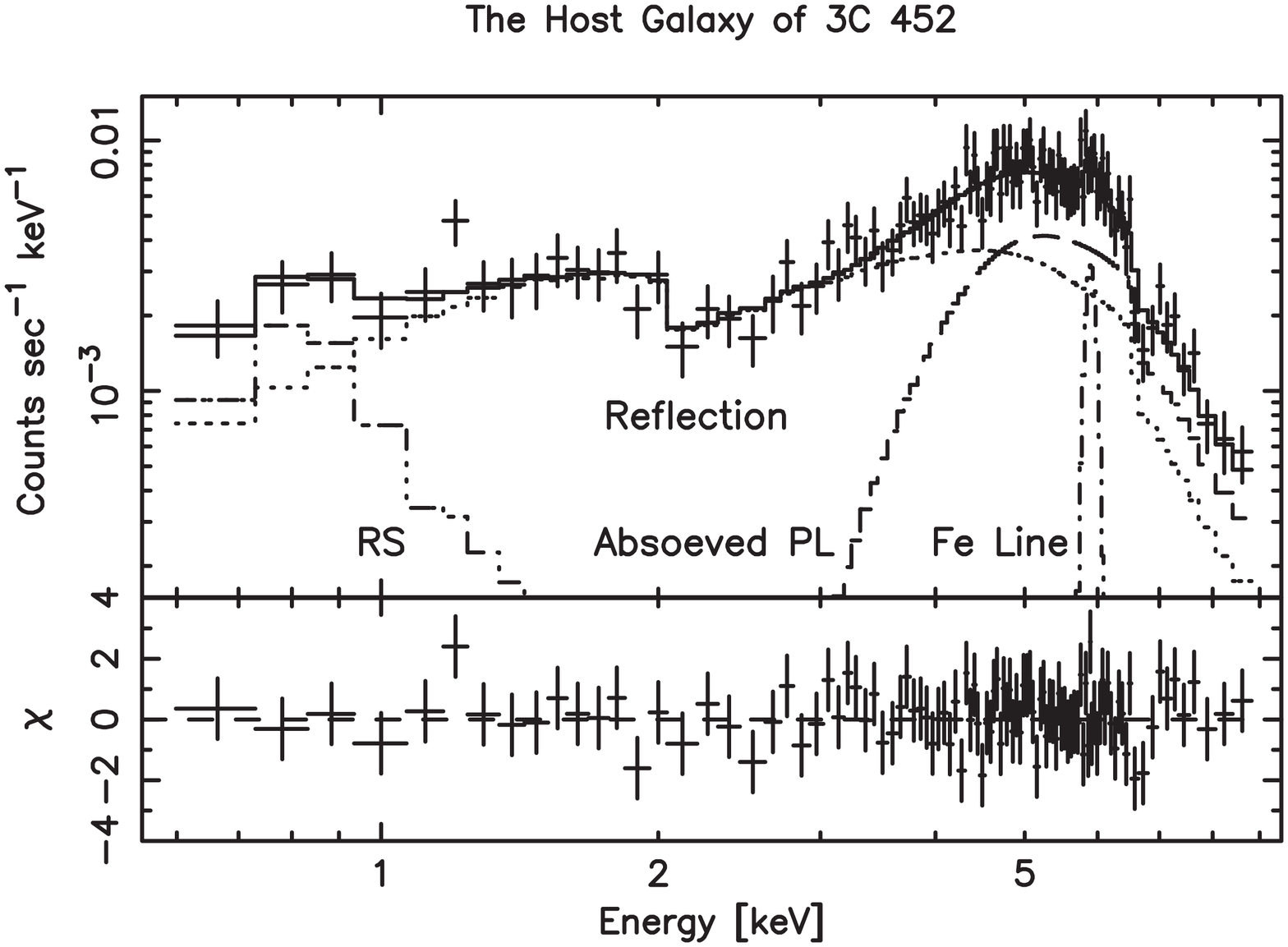}
\caption{The background-subtracted ACIS spectrum of the host galaxy of 3C~452
in the 0.5 -- 9 keV range, 
shown without removing the instrumental response.
The best-fit model consisting of 4 spectral components 
is shown with the histograms.} 
\label{fig:nucl_spec}
\end{figure}

\begin{figure}[htbp]
\plotone{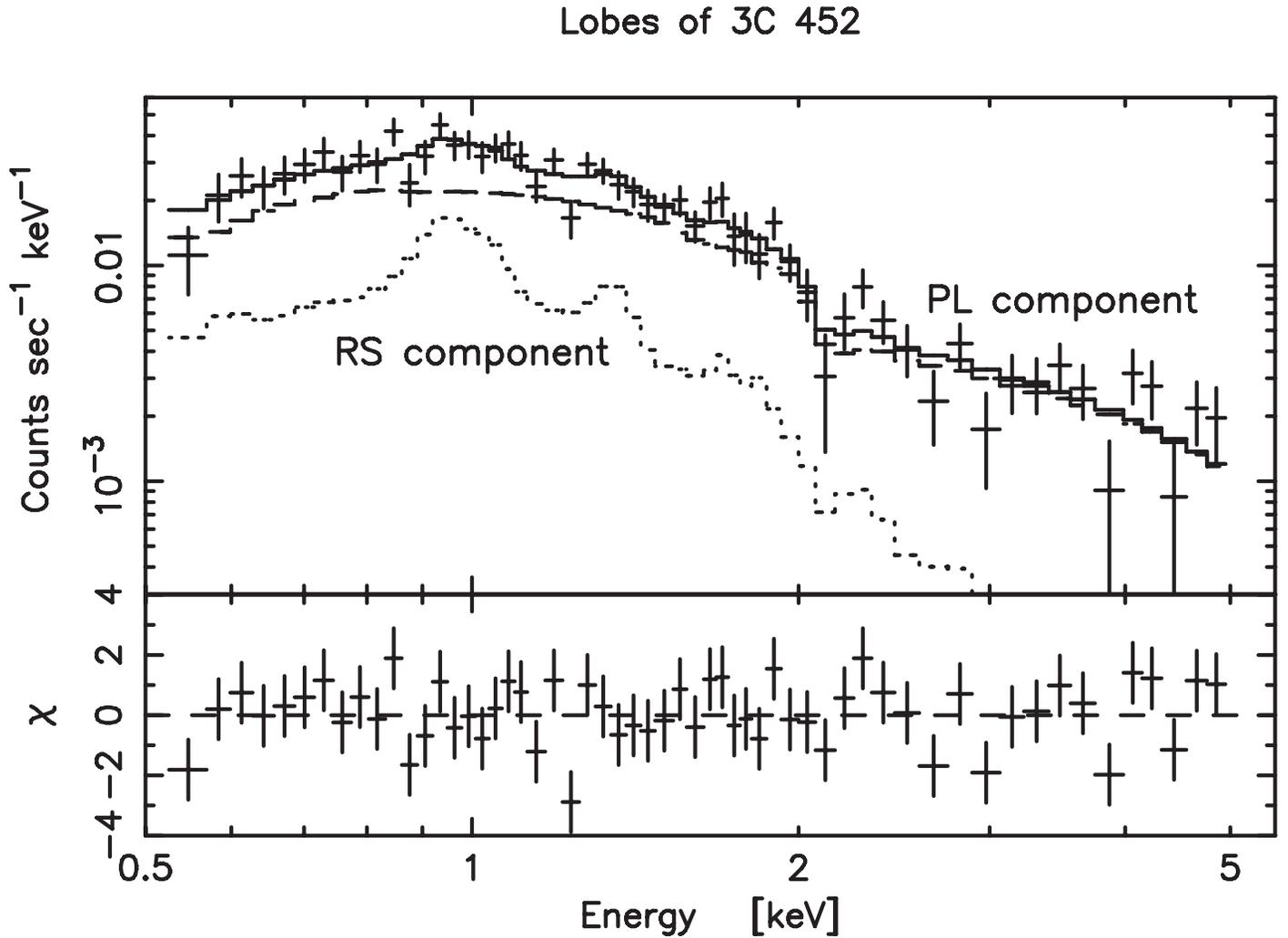}
\caption{The background-subtracted ACIS spectrum of the lobes of 3C~452,
shown without removing the instrumental response.
The dashed and dotted histograms represent the best-fit 
soft RS and hard PL component, respectively,
and the solid one shows their sum.}
\label{fig:lobe_spec}
\end{figure}

Figure 2 visualizes the difference between the SR and IC surface brightness
profiles,
suggesting a significant difference in spatial distributions
between the electrons and magnetic fields.
To reproduce the relatively flat IC X-ray profile,
we first employed a simple model
that the electrons uniformly fill an ellipsoid centered on the nucleus.
As shown in Figure \ref{fig:profile} (a),
the model prediction agrees with the observed profile 
($\chi^2/d.o.f.=1.06$, neglecting the point source positions),
when the major axis of the ellipsoid becomes $2'.2 \pm 0'.1$.
Here, the contribution to the observed profile 
from the soft RS component is less than $\sim 15$\%.
By comparing the observed IC X-ray profile and the model one, 
we next calculated the distribution of \ue~
along the major axis of the lobes, 
and show it in Figure \ref{fig:profile} (c).
We hence conclude that 
the electrons are relatively uniformly distributed over the lobes. 

As shown in Figure \ref{fig:profile} (d),
we have also estimated the magnetic field distribution,
using the ratio of the SR radio brightness 
to the IC X-ray brightness.
Thus, the magnetic field is $B\lesssim1~\mu$G near the nucleus, 
while it increases outward, reaching $B \sim 3 ~\mu$G at a distance of 
$\sim2'$ ($\sim160$ kpc) from the nucleus.
Figure \ref{fig:profile} (e) represents 
the distribution of the ratio $u_{\rm e} / u_{\rm m}$.
Near the center of the lobes,
\ue~thus dominates \um~by more than one order of magnitude,
while  $u_{\rm e}/u_{\rm m}$ decreases toward the lobe edges,   
and becomes $u_{\rm e} / u_{\rm m} = 2 \sim 5$ 
at $\sim2'$ from the nucleus.

The present results agree in three specific points
with the previous detections of IC X-rays from, e.g.,  
Fornax A (Tashiro et al. 2001)  and Centaurus B (Tashiro et al. 1998).
One is that $u_{\rm e}$ usually dominate $u_{\rm m}$,
by an order of magnitude (Isobe 2002).
Another is that the electrons are rather uniformly 
distributed over the lobes.
The other is that the magnetic field becomes stronger 
toward the lobe periphery, probably because of compression.
Thus, these three may be common properties of the radio lobes in general.

\acknowledgments
The unpublished VLA image of 3C~452 (Laing)
is downloaded from 
``An Atlas of DRAGNs''  (http://www.jb.man.ac.uk/atlas), 
by Leahy, Bridle, \& Strom.      
A part of this work was supported by the Grants-in-Aid by the Ministry
of Education, Culture, Sports, Science,  
and Technology (MEXT) of Japan (11440074).


\begin{table}[htbp]
\caption{The best-fit spectral parameters for the host galaxy of
3C~452.}
\label{table:nucle}
\begin{center}
\begin{tabular}{l c l c l c l c }
\hline \hline 
\multicolumn{2}{c}{Absorbed PL}					& \multicolumn{2}{c}{Gaussian}		&\multicolumn{2}{c}{Reflection}	&\multicolumn{2}{c}{RS} \\
\hline 
$N_{\rm H}$(\colden)	& $(5.9\pm0.9)\times10^{23}$		& $E$ (keV)		& $6.40_{-0.7}^{+1.0}$	& \multicolumn{2}{c}{--} 	&  $kT$ (keV)		&$0.79_{-0.16}^{+0.10}$\\
$\alpha $		& $0.66_{-0.21}^{+0.28}$		& $\sigma$ (keV)	& $\lesssim 0.22$	& \multicolumn{2}{c}{--}	&	\multicolumn{2}{c}{--} \\
\hline 
$L_{\rm X}$\tablenotemark{a} &$6.7_{-1.9}^{+1.5}\times10^{43}$	& $EW$ (keV)	\tablenotemark{c}	&$0.32_{-0.20}^{+0.31}$	& $L_{\rm X}$\tablenotemark{a}			&$7.6_{-1.6}^{+2.0}\times10^{42}$ & $L_{\rm X}$ \tablenotemark{b}	& $4.6_{-2.5}^{+3.3}\times10^{40}$	\\
\hline 
\end{tabular}
\tablenotetext{a}{absorption corrected 2 -- 10 keV luminosity in \lum} 
\tablenotetext{b}{absorption corrected 0.5 -- 10 keV luminosity in \lum} 
\tablenotetext{c}{equivalent width against the reflection component} 
\end{center}
\end{table}

\begin{table}[htbp]
\caption{The best-fit spectral parameters of the lobes of 3C~452.}
\label{table:lobe_param}
\begin{center}
\begin{tabular}{lc cc c c}
\hline\hline 
                &                       & \multicolumn{2}{c}{Hard Component}            & Soft Component                &       \\
 Model		& $N_{\rm H}~(10^{21}$ cm$^{-2})$\   & $\alpha_{\rm X}$,  $kT$ (keV) & $S_{\rm X}$ (nJy) \tablenotemark{a} 	& $kT$ (keV)                    & $\chi^2/d.o.f$\\
\hline 
PL              & $2.1\pm0.5$           & $1.02_{~-0.19}^{~+0.21}$& $66\pm 10$   & --                            & $72.8/54$     \\
PL + RS         & $1.6_{~-0.4}^{~+0.5}$ & $0.68\pm{0.28}$       & $41 \pm 15$    & $1.36_{~-0.29}^{~+0.47}$      & $62.0/52$     \\
Bremss          & $1.3_{~-0.3}^{~+0.4}$ & $3.7_{~-1.0}^{~+1.3}$ &       --              & --                            & $77.6/54$\\
Bremss + RS     & $1.4_{~-0.4}^{~+0.3}$ & $6.0_{~-1.7}^{~+13.5}$&       --              & $1.28_{~-0.17}^{~+0.29}$      & $63.7/52$     \\
\hline 
\end{tabular}
\tablenotetext{a}{flux density at 1 keV} 
\end{center}
\end{table}

\end{document}